\documentclass[%
 reprint,
 amsmath,amssymb,
 aps,
]{revtex4-1}

\usepackage{dcolumn}
\usepackage{bm}

\usepackage{ascmac}

\usepackage{graphicx}
\usepackage{color}

\newcommand{\eq}[1]{\begin{equation} #1 \end{equation}}
\newcommand{\eqa}[2]{\begin{equation} #1 \label{#2} \end{equation}}
\newcommand{\balign}[1]{\begin{align} #1 \end{align}}
\newcommand{\bcases}[1]{\begin{cases} #1 \end{cases}}

\DeclareMathOperator*{\argmax}{arg\,max}

\newcommand{\figin}[4]
{\begin{figure}[tb]
\centering
\includegraphics[width= #1]{#2.pdf}
\caption{#3}
\label{f:#4}
\end{figure}}

\newcommand{\todayd}{\the\year/\the\month/\the\day}
\newcommand{\del}{\partial}
\newcommand{\bib}{\bibitem}

\newcommand{\lr}{\leftrightarrow}
\newcommand{\const}{\mathrm{const}}

\newcommand{\lb}{\label}
\newcommand{\nt}{\notag}

\newcommand{\ft}[2]{\left. #1 \right|_{#2}}

\newcommand{\eref}[1]{Eq.~\eqref{#1}}

\newcommand{\fref}[1]{Fig.~\ref{f:#1}}

\newcommand{\bel}{\begin{easylist}}
\newcommand{\eel}{\end{easylist}}
\newcommand{\bi}[1]{\begin{itemize} #1 \end{itemize}}
\newcommand{\be}[1]{\begin{enumerate} #1 \end{enumerate}}

\def \({\left(}
\def \){\right)}
\def \[{\left[}
\def \]{\right]}
\newcommand{\la}{\left\langle}
\newcommand{\ra}{\right\rangle}

\newcommand{\sumtwo}[2]%
{\mathop{\sum_{#1}}_{#2}}
\newcommand{\sumthree}[3]%
{\mathop{\mathop{\sum_{#1}}_{#2}}_{#3}}
\newcommand{\sumfour}[4]%
{\mathop{\mathop{\mathop{\sum_{#1}}_{#2}}_{#3}}_{#4}} 
\newcommand{\prodtwo}[2]%
{\mathop{\prod_{#1}}_{#2}}
\newcommand{\mintwo}[2]%
{\mathop{\min_{#1}}_{#2}}
\newcommand{\maxtwo}[2]%
{\mathop{\max_{#1}}_{#2}}
\newcommand{\maxthree}[3]%
{\mathop{\mathop{\max_{#1}}_{#2}}_{#3}}
\newcommand{\limtwo}[2]%
{\mathop{\lim_{#1}}_{#2}}
\newcommand{\suptwo}[2]%
{\mathop{\sup_{#1}}_{#2}}
\newcommand{\supthree}[3]%
{\mathop{\mathop{\sup_{#1}}_{#2}}_{#3}}
\newcommand{\supfour}[4]%
{\mathop{\mathop{\mathop{\sup_{#1}}_{#2}}_{#3}}_{#4}} 
\newcommand{\inftwo}[2]%
{\mathop{\inf_{#1}}_{#2}}
\newcommand{\infthree}[3]%
{\mathop{\mathop{\inf_{#1}}_{#2}}_{#3}}
\newcommand{\inffour}[4]%
{\mathop{\mathop{\mathop{\inf_{#1}}_{#2}}_{#3}}_{#4}} 

\newcommand{\ep}{\varepsilon}

\newcommand{\Di}{\mathit{\Delta}}

\newcommand{\muH}{\mu_{\rm H}}
\newcommand{\muL}{\mu_{\rm L}}
\newcommand{\nuH}{\nu^*_{\rm H}}
\newcommand{\nuL}{\nu^*_{\rm L}}
\newcommand{\hJ}{\hat{J}}

\def\rnum#1{\resizebox{0.5em}{\height}{\expandafter{\romannumeral #1}}}
\def\Rnum#1{\resizebox{0.5em}{\height}{\uppercase\expandafter{\romannumeral #1}}}

\begin{document}

\preprint{APS/123-QED}

\title{Stationary engines in and beyond the linear response regime at the Carnot efficiency}

\author{Naoto Shiraishi}
\affiliation{Department of Physics, Keio University, 3-14-1 Hiyoshi, Yokohama 223-8522, Japan} %
\date{\today}

\begin{abstract}
The condition for stationary engines to attain the Carnot efficiency in and beyond the linear response regime is investigated.
We find that this condition for finite-size engines is significantly different from that for macroscopic engines in the thermodynamic limit.
For the case of finite-size engines, the tight-coupling condition in the linear response regime directly implies the attainability of the Carnot efficiency beyond the linear response regime.
Contrary to this, for the case of macroscopic engines in the thermodynamic limit, there are three types of mechanisms to attain the Carnot efficiency.
One mechanism allows engines to attain the Carnot efficiency only in the linear response limit, while other two mechanisms enable engines to attain the Carnot efficiency beyond the linear response regime.
These three mechanisms are classified by introducing tight-coupling window.

\begin{description}
\item[PACS numbers]
05.70.Ln, 05.40.-a, 87.10.Mn, 87.16.Nn.
\end{description}
\end{abstract}

\pacs{Valid PACS appear here}
\maketitle

\section{Introduction}

Nonequilibrium stationary systems with stationary currents are widely-used setups to investigate nonequilibrium phenomena in statistical mechanics.
Stationary systems and stationary currents (e.g., heat current, electric current, and matter current) are ubiquitous from biochemical systems to engineering products.
If two different types of currents flow in a single system, this system can behave as an engine in the following sense:
One current flows obeying the conjugate thermodynamic force, and it induces another current flowing against the conjugate thermodynamic force.
In the case of the coupling of heat current and electric current, this phenomenon is known as the Seebeck effect and the Peltier effect.
Molecular motors in living systems are also stationary engines which convert chemical potential consumption to mechanical work or another particle current.

The framework of nonequilibrium statistical mechanics for stationary systems in the linear response regime is well-established~\cite{Todabook}.
On the basis of this framework, stationary engines in the linear response regime are well-studied and their ability is evaluated.
The most important quantity of engines is efficiency, whose maximum is known as the Carnot efficiency (CE).
It is established that the general necessary and sufficient condition for stationary engines in the linear response regime to attain the CE is the tight-coupling condition (i.e., the determinant of the Onsager matrix is zero)~\cite{Bro07}.

Contrary to the case in the linear response regime, general results on stationary engines beyond the linear response regime have been missing.
Most studies on such engines have analyzed specific elaborated models including Feynman's ratchet~\cite{Feynman, PE, Sekimoto97, JM}, the Brownian motors~\cite{But, Lan, HS, Matsuo, DA}, the sensor-gate model or the autonomous Maxwell's demon~\cite{Sek05, Esposito, SS}, ideal diodes~\cite{Sok}, soft nanomachines~\cite{Seifert}, and mesoscopic thermoelectric conductors~\cite{Hum, MS, SB, HPC}, and some of the proposed models attain the CE ~\cite{MS, Sok, DA, Hum, HPC, Seifert, Sek05, Esposito, SS, SB}.
However, general properties for stationary engines are only suggested in some papers that a kind of singularity might be important to attain the CE beyond the linear response regime~\cite{Sekimotobook, MS}.
There are also some attempts to characterize tight-coupling beyond the linear response regime~\cite{EKLB, Wang}, while the connection to the CE has not been revealed.

In this paper, we investigate the condition for stationary engines to attain the CE.
By employing the method of partial entropy production~\cite{SS} and the zeroth law of thermodynamics, we derive the necessary condition for general stationary engines to attain the CE.
Our derivation follows the spirit in Ref.~\cite{Shi15}, while ours is more sophisticated and covers broader classes of stationary engines.
We then clarify the difference between the case of finite-size engines and that of macroscopic engines in the thermodynamic limit.
We find that for the case of finite-size engines the condition to attain the CE in the linear response regime (i.e., tight-coupling condition in conventional sense) and that beyond the linear response regime are completely same.
In contrast, for the case of macroscopic engines in the thermodynamic limit, there are three different types of engines which attain the CE.
The first type attains the CE only in the linear response limit, and the second type can always attain the CE regardless of the amount of the chemical potential difference.
The third type can attain the CE with a pair of chemical potentials within a certain definite range.
The second and the third ones employ different types of singularity to realize the CE.
Our aforementioned findings are well characterized by the tight-coupling window, which manifests how and what types of tight-coupling appears beyond the linear response regime.

This paper is organized as follows.
Before going to general discussion, in Sec.~\ref{models}, we introduce two models of stationary engines which serve respectively as prototypes of finite-size stationary engines and macroscopic stationary engines.
These models attain the CE with some choice of parameters, while they do not with some other choice.
The main part of this paper is Sec.~\ref{gen}, in which we clarify the general necessary and sufficient condition to attain the CE.
In Sec.~\ref{pep}, we explain the method of partial entropy production, and using this, we demonstrate an important property which should be satisfied in engines at the CE.
With the aid of this properties, in Sec.~\ref{gen-pri}, we derive the necessary and sufficient condition to attain the CE.
In Sec.~\ref{fin} and Sec.~\ref{macro}, we show some consequences of this condition for finite-size engines and for macroscopic engines, respectively.
In Sec.~\ref{window}, we classify these findings by introducing tight-coupling window.

In the most part of this paper, we consider isothermal engines driven by two particle baths with different chemical potentials, since the extension of our results to the case of heat engines driven by two heat baths is straightforward.
We remark that the maximum of the efficiency for such engines (i.e., work extraction divided by chemical potential consumption) is not $1-T_{\rm L}/T_{\rm H}$ but 1, which we also call as the Carnot efficiency.

\section{Two models of stationary engines}\lb{models}

Before going to general discussion, we first introduce two models of stationary engines, the information engine with switch and the coarse-grained autonomous Carnot engine, which are prototypes of finite-size and macroscopic stationary engines, respectively.

\subsection{Information engine with switch}\lb{IES}

We first introduce the information engine with switch (IES), in which a wall with mechanical force and particles driven by chemical potential difference affect each other in the form of information.
The IES consists of three parts; a site,  a wall,  and a switch, which are attached to a single heat bath with inverse temperature $\beta$ (see \fref{add-info}).
The site can store at most one particle, where the number of the particle is denoted by $n\in\{0,1\}$.
Two particle baths H and L with chemical potentials $\muH$ and $\muL$ ($\muH>\muL$)  exchange particles only with this site.
The wall places between the site and one of the baths.
The position of the wall is denoted by $x\in\{l,r\}$. 
If $x=l$ ($x=r$), the wall prohibits the jump of particles between the site and the bath H (L). 
The switch $e\in \{ e_1,e_2\}$ changes the coefficients of transition rates of the wall.
The state of the whole system is described by $(x,n,e)$.

\figin{7cm}{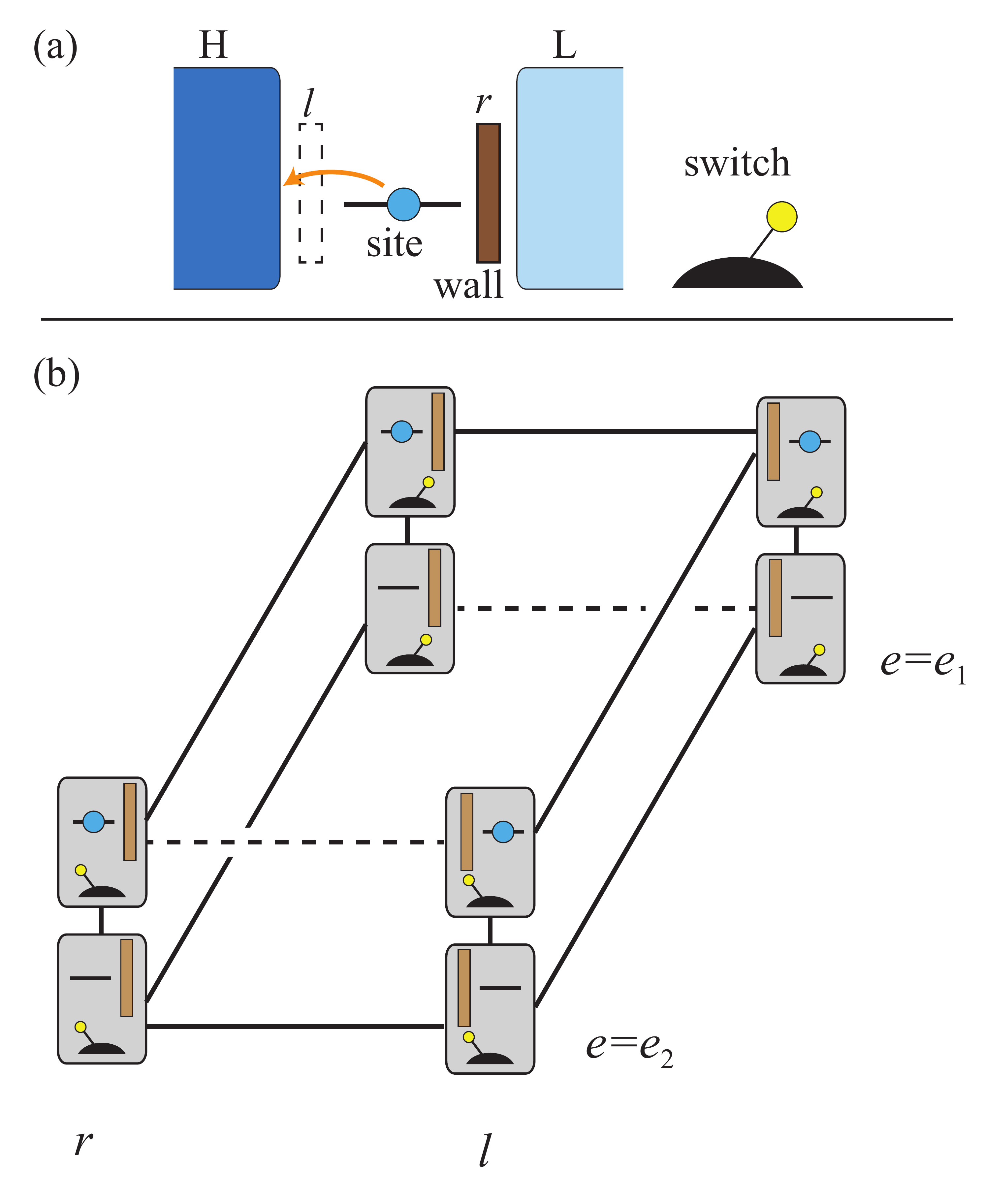}{
(a): Schematic of the information engine with switch (IES).
A site lies in the middle of two particle baths H and L, and a wall at $r$ ($l$) prevent the jump of particles between the site and the bath L (H).
The switch modulates the transition rates of the wall.
(b): State space of the IES.
The dashed transition paths vanish if we set $\ep=0$.
}{add-info}

The transition rates of $e$ is set as a constant independent of $x$ and $n$:
\eq{
P_{(x,n,e_1)\to (x,n,e_2)}=P_{(x,n,e_2)\to (x,n,e_1)}=c.
}
The transition rates of $x$ and $n$ are set as
\balign{
P_{(l,0,e_1)\to (l,1,e_1)}=P_{(l,0,e_2)\to (l,1,e_2)}=&P_{(l,0)\to (l,1)}, \\
P_{(l,1,e_1)\to (l,0,e_1)}=P_{(l,1,e_2)\to (l,0,e_2)}=&P_{(l,1)\to (l,0)}, \\
P_{(r,0,e_1)\to (r,1,e_1)}=P_{(r,0,e_2)\to (r,1,e_2)}=&P_{(r,0)\to (r,1)}, \\
P_{(r,1,e_1)\to (r,0,e_1)}=P_{(r,1,e_2)\to (r,0,e_2)}=&P_{(r,1)\to (r,0)}, 
}
and
\balign{
P_{(r,1,e_1)\to (l,1,e_1)}=&(2-\ep)P_{r\to l}^-, \\
P_{(l,1,e_1)\to (r,1,e_1)}=&(2-\ep)P_{l\to r}^-, \\
P_{(r,1,e_2)\to (l,1,e_2)}=&\ep P_{r\to l}^+, \\
P_{(l,1,e_2)\to (r,1,e_2)}=&\ep P_{l\to r}^+, \\
P_{(r,0,e_1)\to (l,0,e_1)}=&\ep P_{r\to l}^-, \\
P_{(l,0,e_1)\to (r,0,e_1)}=&\ep P_{l\to r}^-, \\
P_{(r,0,e_2)\to (l,0,e_2)}=&(2-\ep)P_{r\to l}^+, \\
P_{(l,0,e_2)\to (r,0,e_2)}=&(2-\ep)P_{l\to r}^+, 
}
with $0\leq \ep\leq 1$.
The transition rates of the particles satisfy the following detailed-balance condition~\cite{LDB}:
\balign{
\ln \frac{P_{(r,0)\to (r,1)}}{P_{(r,1)\to (r,0)}}=&\beta\muH, \lb{4state1} \\
\ln \frac{P_{(l,0)\to (l,1)}}{P_{(l,1)\to (l,0)}}=&\beta\muL, \lb{4state2}
}
and those of the wall also satisfy
\balign{
\ln \frac{P_{r\to l}^+}{P_{l\to r}^+}=&\beta E^+, \lb{4state3}\\
\ln \frac{P_{r\to l}^-}{P_{l\to r}^-}=&\beta E^-. \lb{4state4}
}
Here, $E^+$ and $E^-$ are energy differences applied externally to the wall, which we in particular set as $E^+>E^-$.
In this model, no direct energy exchange between the particles and the wall occurs, and the particle flow and the replacement of the wall affect each other only through the correlation (information) between the presence/absence of the particle in the site and the position of the wall.
The switch changes its own state completely random, and it modulates the transition rates as in the following manner:
When $e=e_1$ ($e=e_2$), the wall is likely to change its state if a particle is (is not) in the site.
We note that this additional switch is different from the additional variable introduced in Ref.~\cite{SIKS}.
If $\muH-\muL>E^+-E^-$ and $\ep$ is sufficiently small, the particles are transported from the bath H to L and the wall performs work against the external force.

By taking $c\to \infty$ limit and coarse-graining the fast variable $e$, this model reduces to a model known as the sensor-gate model~\cite{Sek05} or the autonomous Maxwell's demon~\cite{Esposito, SS, SIKS, YISS, AJM, HBS, SMS, HE}.
For the case with $\ep=0$, the reduced system is tight-coupling in the conventional sense (i.e., the determinant of the Onsager matrix is zero), and it attains the CE both in and beyond the linear response regime.
In contrast, for the case with $\ep>0$, inevitable particle leakage exists and it never attains the CE.

\subsection{Coarse-grained autonomous Carnot engine}\lb{CGACE}

We next introduce the coarse-grained autonomous Carnot engine (CGACE)~\cite{Shi15}, which is an autonomous version of the engine driven by two particle baths with chemical potentials $\muH $ and $\muL $ denoted by H and L (see \fref{aC-cycle}).
The engine is in the isothermal condition with inverse temperature $\beta$
The state of the engine denoted by $X$ takes four possible states,  $A$, $B$, $C$, and $D$, whose volumes are $V_A$, $V_B$, $V_C$ and $V_D$.
The former two states are attached to the bath H, and the latter two states are attached to the bath L.
We here define  the number of particles divided by $V_A$ as $\nu$ in order to consider the thermodynamic limit $V_A\to \infty$ finally.
The process $A\lr B$ ($C\lr D$) is associated with the bath H (L), while the processes $B\lr C$ and $A\lr D$ conserve the number of particles, which we call closed processes.
The change in states is described by a Markov jump process.
We consider the case that the engine tends to move along $A\to B\to C\to D\to A$ due to the particle current from the bath H to L, and we extract the work by imposing external force in the direction $A\to D\to C\to B\to A$.
We here suppose that the two closed processes are slow processes and the probability distribution for states with $A$ and $B$ ($C$ and $D$) is given by the grand canonical distribution with $\muH$ ($\muL$).

We define $E_{AB}$ as the energy difference from $A$ to $B$ including the external force, and $E_{BC}$, $E_{CD}$, $E_{DA}$ are defined in a similar manner.
With noting that the total change in potential energy is zero through a cyclic process, the work done by the engine against the external force through a single rotation $A\to B\to C\to D\to A$ is expressed as 
\eq{
E_{AB}+E_{BC}+E_{CD}+E_{DA}=:W_{\rm tot}>0.
}
The isothermal condition of the engine implies that the transition rates of $A\lr D$ and $B\lr C$ denoted by $P_{X\to X^-;\nu V_A}$ satisfy the following detailed-balance condition:
\eqa{
\ln \frac{P_{X\to X^-;\nu V_A}}{P_{X^-\to X;\nu V_A}}=-\beta (e_{XX^-}+F(r_{X^-},\nu)-F(r_X,\nu))V_A,
}{ldb}
where the superscript $-$ represents the state of its opposite side: $A^-:=D, B^-:=C, C^-:=B, D^-:=A$, and $F(V,n)$ is the Helmholtz free energy.
We defined normalized energy and the ratio of $V_A$ to $V_X$ as
\balign{
e_{XX^-}&:=-e_{X^-X}=\frac{E_{XX^-}}{V_A}, \\
r_X&:=\frac{V_X}{V_A}.
}
The extensive property of the free energy implies $F(r,\nu)=F(V,n)/V_A$.

\figin{8cm}{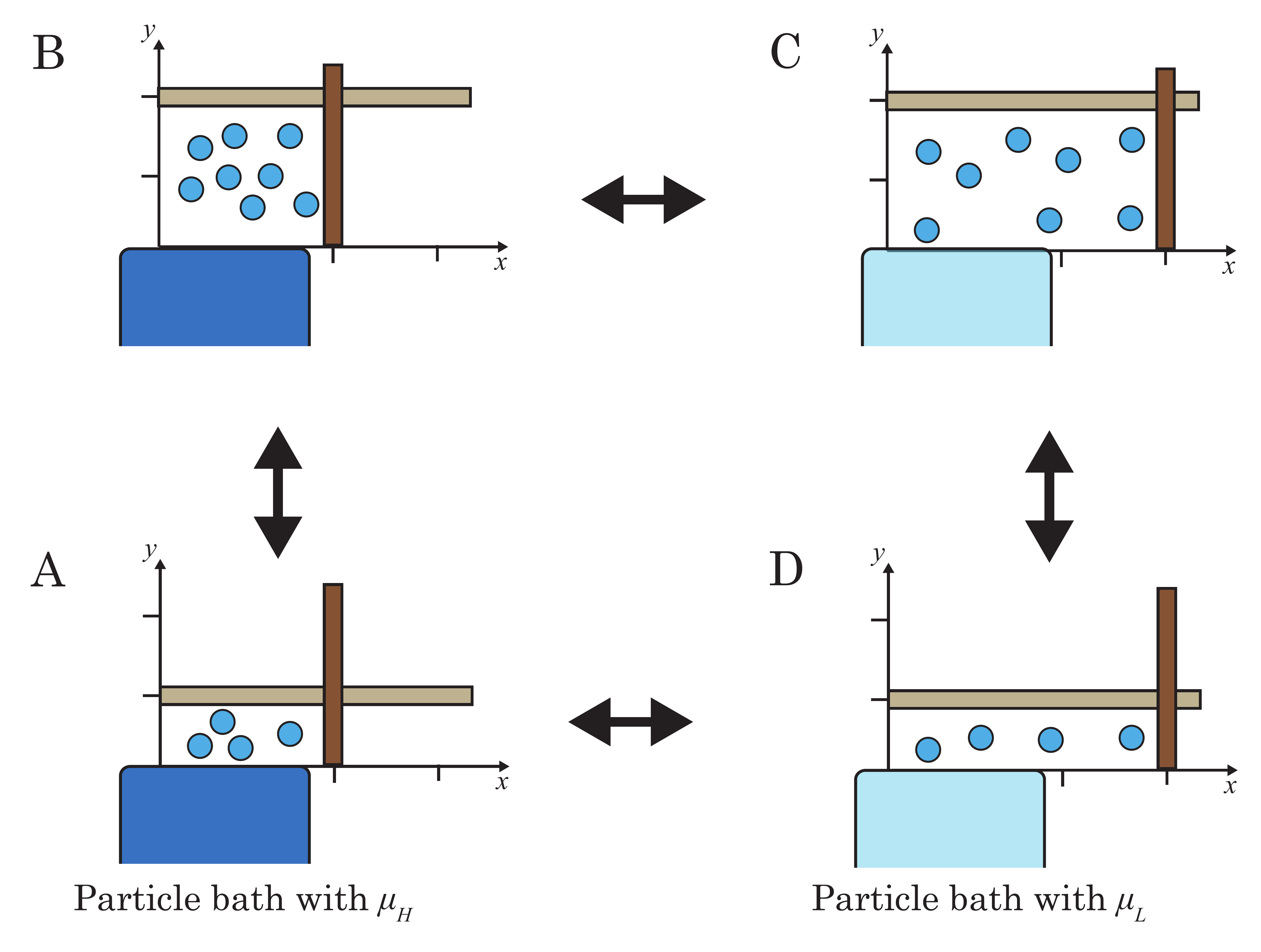}{
State space of the coarse-grained autonomous Carnot engine (CGACE).
The transitions with $B\lr C$ and $A\lr D$ are isothermal but not attached to particle baths.
Work is extracted via the rotational path $A\to B\to C\to D\to A$.
Although the engine contains few particles in this figure, it actually contains infinitely many particles.
}{aC-cycle}

We now consider the situation of large $V_A$.
The stationary distribution of probability density with respect to $\nu$ is calculated as
\balign{
P^{\rm ss}(A,\nu)&=V_A P^{\rm ss}_{AB}\frac{e^{-\beta (F(r_A,\nu)-\muH \nu )V_A}}{Z_{AB}}, \\
P^{\rm ss}(B,\nu)&=V_A P^{\rm ss}_{AB}\frac{e^{-\beta (F(r_B,\nu)+e_{AB}-\muH \nu) V_A}}{Z_{AB}}, 
}
where $P^{\rm ss}_{AB}$ represents the stationary probability at $A$ or $B$, and 
\balign{
Z_{AB}:=\sum_\nu&e^{-\beta (F(r_A,\nu)-\muH \nu )V_A} \nt \\
&+e^{-\beta (F(r_B,\nu)+e_{AB}-\muH \nu) V_A}
}
is a normalization constant ensuring
\eq{
\int d\nu [P^{\rm ss}(A,\nu)+P^{\rm ss}(B,\nu)]=1.
}
We define $P^{\rm ss}(C,\nu)$, $P^{\rm ss}(D,\nu)$, and $Z_{CD}$ in a similar manner.
The stationary probability flow of $X\to X^-$ with $\nu$ is given by 
\eqa{
j_{X\to X^-}(\nu):=P^{\rm ss}(X,\nu)P_{X\to X^-;\nu V_A}.
}{jXX-}
The law of large numbers suggests that the realized number of particles under the condition that the transition $X\to X^-$ occurs is around the most provable value
\eqa{
\nu_X^*:=\argmax _\nu j_{X\to X^-}(\nu).
}{nX*}
As will be discussed in the Appendix.~\ref{detail-CGACE}, the CGACE does not attain the CE with nonsingular setups, while it attains the CE with some singular setups.

\section{General necessary condition for Carnot efficiency}\lb{gen}

\subsection{Partial entropy production and its implication}\lb{pep}

Before going to the derivation of the necessary and sufficient condition for stationary engines to attain the CE, we first explain the method of partial entropy production and its implication, which is first shown in Ref.~\cite{SS16}.
Consider a Markov jump process on discrete states.
Let $p_w$ and $P_{w\to w'}$ be the probability distribution of $w$ and the transition rate from $w$ to $w'$.
We then define the dual transition rate $\tilde{P}_{w'\to w}$ as satisfying
\eqa{
\ln \frac{P_{w\to w'}}{\tilde{P}_{w'\to w}}=\beta (E_w-E_{w'})-\beta \mu (n_w-n_{w'}),
}{def-dual}
where $E_w$ and $n_w$ are the energy and the number of particles of the state $w$.
The right-hand side of \eref{def-dual} is equal to the thermodynamic force on the transition $w\to w'$.
The entropy production of the total system is then given by~\cite{Sei12}
\eq{
\sigma :=\sum_{w,w'}p_wP_{w\to w'}\ln \frac{p_wP_{w\to w'}}{p_{w'}\tilde{P}_{w'\to w}},
}
where two terms $\sum_{w,w'}p_wP_{w\to w'}\ln ({p_w}/{p_{w'}})$ and $\sum_{w,w'}p_wP_{w\to w'}\ln (P_{w\to w'}/\tilde{P}_{w'\to w})$ correspond to the change in the Shannon entropy of the system and that in the heat dissipation times inverse temperature of the bath, respectively.
Precisely speaking, this quantity is not entropy production but entropy production rate, but we also call this quantity as entropy production in case of no confusion. 
We now introduce a key quantity: the (averaged) partial entropy production with a transition $w\to w'$ as
\balign{
\sigma_{w\to w'}:=&p_wP_{w\to w'}\ln \frac{p_wP_{w\to w'}}{p_{w'}\tilde{P}_{w'\to w}} \nt \\
&+p_{w'}\tilde{P}_{w'\to w}-p_wP_{w\to w'},
}
which is a decomposition of the entropy production into each transition:
\eq{
\sigma =\sum_{w,w'}\sigma_{w\to w'}.
}

The nonnegativity of the partial entropy production is confirmed by a simple mathematical inequality:
\eqa{
a\ln \frac{a}{b}+b-a\geq 0.
}{math-partial}
A similar inequality is used to show the nonnegativity of the relative entropy.
An important consequence of this fact is that zero entropy production ($\sigma=0$) leads to zero partial entropy production for all transitions ($\sigma_{w\to w'}=0$ for all $w\to w'$).
In addition, the equality of the above inequality \eqref{math-partial} holds if and only if $a=b$.
Since the case of  $a=p_wP_{w\to w'}$ and $b=p_{w'}\tilde{P}_{w'\to w}$ corresponds to the partial entropy production, we find that zero partial entropy production $\sigma_{w\to w'}=0$ implies
\eq{
\frac{p_w}{p_{w'}}=\frac{\tilde{P}_{w'\to w}}{P_{w\to w'}}=e^{-\beta (E_w-E_{w'})+\beta \mu (n_w-n_{w'})},
}
which means that the rate of probability distribution between $w$ and $w'$ is equal to corresponding equilibrium distribution.
The aforementioned discussion is summarized as follows:
Zero entropy production is equivalent to zero partial entropy production for any transition $w\to w'$, which directly implies that the ratio of probability distributions with $w$ and $w'$ is equal to corresponding equilibrium distribution.

\subsection{General condition to attain the Carnot efficiency}\lb{gen-pri}

We now derive the necessary and sufficient condition for stationary engines to attain the CE.
We denote the state of the whole system by $(X,n)$ (or $(X,\nu)$ for the case in the thermodynamic limit), where $X$ represents the state of the engine and $n$ represents the number of particles ($\nu$ represents the number of particles divided by a typical volume $V_0$).
For the case of engines driven by two heat baths, the state of the whole system is denoted by $(X,E)$ with the energy $E$.
In the following, we consider a stationary engine at the CE and clarify the conditions this engine should satisfy.

We first ensure that an engine never attains the CE if there exist two states $(X,n)$ and $(X',n')$ ($n\neq n'$) such that the bath H induces a transition $(X,n)\to \cdots \to (X',n')$ and the bath L induces another transition $(X,n)\to \cdots \to (X',n')$~\cite{comment1}, which we call  {\it bath separation condition}.
We shall explain why this condition is required.
As seen in the previous subsection, the entropy production goes to zero only when the local distribution is equilibrium one.
However, two transition paths with different chemical potentials never satisfy the equilibrium condition simultaneously, which implies finite entropy production and less efficiency than the CE.
The bath separation condition implies that the engine is attached to at most a single particle bath at one moment.
We thus decompose the states of the engine $\{ X\}$ into those attached to the bath H expressed as $\{ X_{1,{\rm H}},\cdots ,X_{n,{\rm H}}\}$ and those to the bath L as $\{ X_{1,{\rm L}},\cdots ,X_{m,{\rm L}}\}$.
In the case of the IES, the engine takes four possible states $(r,e_1)$, $(l,e_1)$, $(r,e_2)$, and $(l,e_2)$, which are decomposed into $\{ (r,e_1), (r,e_2) \} =\{ X_{1,{\rm H}},X_{2,{\rm H}}\}$ and $\{ (l,e_1), (l,e_2)\} =\{ X_{1,{\rm L}}, X_{2,{\rm L}}\}$.
In the case of the CGACE, the engine takes four possible states $\{ A,B,C,D\}$, which are decomposed into  $\{ A,B\} =\{ X_{1,{\rm H}},X_{2,{\rm H}}\}$ and $\{ C,D\} =\{ X_{1,{\rm L}}, X_{2,{\rm L}}\}$.

We suppose that the energy difference of the engine between two states (e.g., $E_{XX^-}$ in the CGACE) is independent of $n$.
Two models introduced in the previous section satisfy this condition.
If a model does not satisfy this condition, we modify the model by adding a new parameter to satisfy this condition (details are discussed in the Appendix.~\ref{add-para}).
This setup ensures that the retracing between $X$ and $X^-$ does not contribute to the amount of work extraction.

Employing the result in the previous subsection again, we further obtain that the stationary distribution of $n$ with the state of the engine $X$ satisfies $P^{\rm ss}(n|X)\propto e^{\beta \mu n}$ if the engine is at the CE.
We call this condition as {\it equilibration condition}~\cite{comment2}.

We now focus on a transition $X_{j,{\rm L}}\to X_{i,{\rm H}}$.
Using the bath separation condition and equilibration condition, the average number of particles transported from $X_{j,{\rm L}}$ to $X_{i,{\rm H}}$ under the condition that the transition from $X_{j,{\rm L}}$ to $X_{i,{\rm H}}$ occurs is calculated as
\eqa{
\la n\ra _{X_{j,{\rm L}}\to X_{i,{\rm H}}} = \sum _n\frac{n\cdot g(n)}{\sum_{n'}g(n')},
}{LtoH}
where $g(n)$ is defined as
\eqa{
g(n):=\frac{e^{-\beta (F(X_{j,{\rm L}},n)-\muL n)}}{\sum_{n'} e^{-\beta (F(X_{j,{\rm L}},n')-\muL n')}} P_{X_{j,{\rm L}}\to X_{i,{\rm H}};n}.
}{auto-def-gn}
We next calculate $\la n\ra _{X_{i,{\rm H}}\to X_{j,{\rm L}}}$.
To calculate this, let us consider an imaginary situation that the chemical potential of the bath L is not $\muL$ but $\muH$ and the whole system is in equilibrium (see \fref{imaginary}).
We express quantities in this imaginary setup by labeling the superscript ``im".
The zeroth law of thermodynamics claims that there is no particle current between $X_{j,{\rm L}}$ and $X_{i,{\rm H}}$ in the imaginary setup:
\eq{
\la n\ra _{X_{j,{\rm L}}\to X_{i,{\rm H}}}^{\rm im}=\la n\ra _{X_{i,{\rm H}}\to X_{j,{\rm L}}}^{\rm im}.
}
In addition, since the distribution of particles in $X_{i,{\rm H}}$ is same in the imaginary setup and in the actual stationary engine, we find
\eq{
\la n\ra _{X_{i,{\rm H}}\to X_{j,{\rm L}}}^{\rm im}=\la n\ra _{X_{i,{\rm H}}\to X_{j,{\rm L}}}.
}
Combining them, we arrive at the expression of $\la n\ra _{X_{i,{\rm H}}\to X_{j,{\rm L}}}$ in terms of $g(n)$ as
\eqa{
\la n\ra _{X_{i,{\rm H}}\to X_{j,{\rm L}}} =\la n\ra _{X_{j,{\rm L}}\to X_{i,{\rm H}}}^{\rm im}= \sum _n\frac{n\cdot g(n)e^{\beta n\Di\mu}}{\sum_{n'}g(n') e^{\beta n'\Di \mu}},
}{HtoL}
where we defined $\Di \mu:=\muH-\muL$.

\figin{7.5cm}{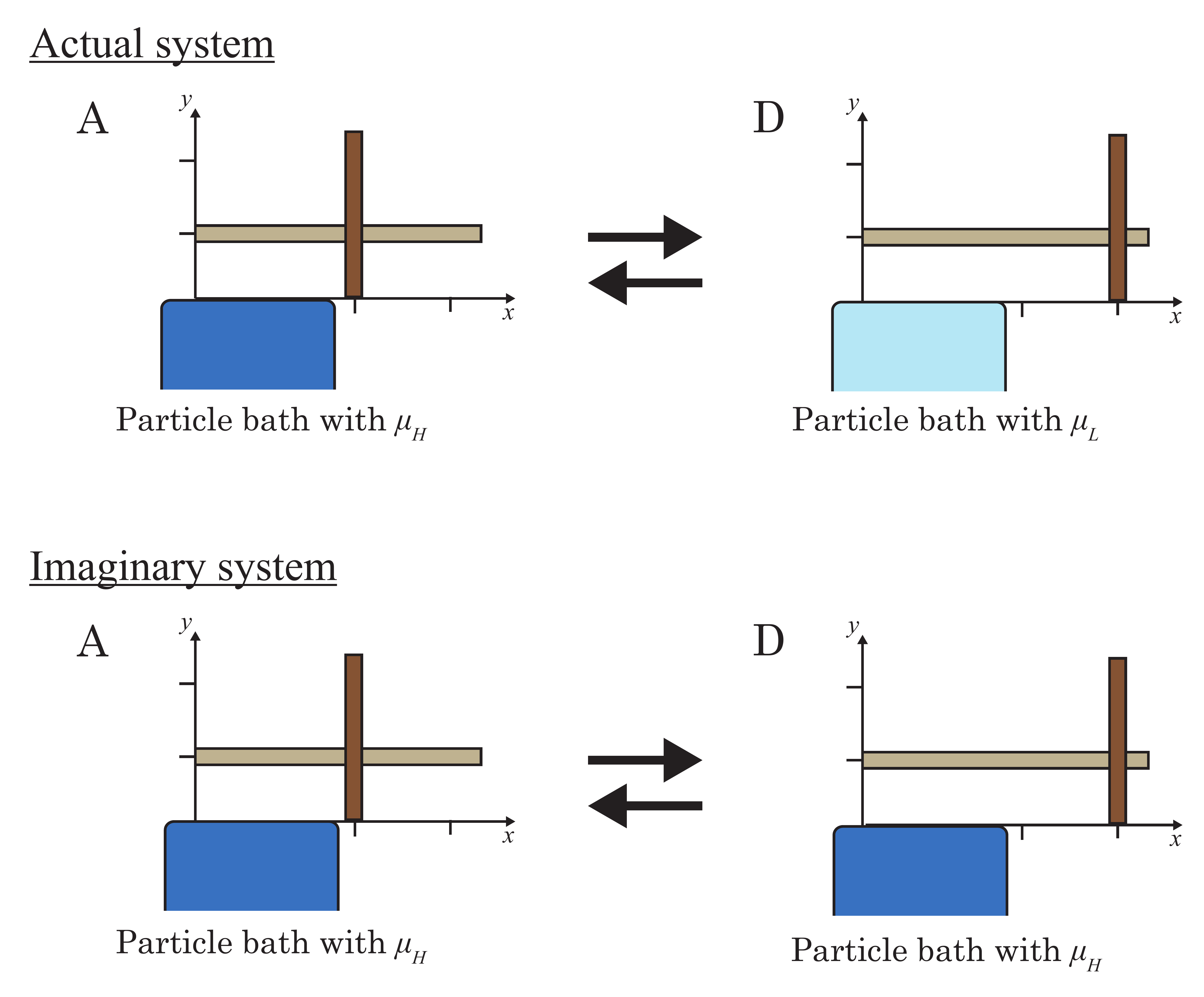}{
Schematic picture of the {\it imaginary system} for the case of the CGACE.
In the imaginary system, the chemical potential of the particle bath attached to the state $D$ is replaced from $\muL$ to $\muH$.
}{imaginary}

The necessary condition for absence of particle leakage (i.e., finite entropy production) between $X_{i,{\rm H}}$ and $X_{j,{\rm L}}$ is expressed as
\eqa{
\la n\ra _{X_{i,{\rm H}}\to X_{j,{\rm L}}}-\la n\ra _{X_{j,{\rm L}}\to X_{i,{\rm H}}} =0
}{zero-leak}
for finite-size engines and 
\eq{
\la \nu \ra _{X_{i,{\rm H}}\to X_{j,{\rm L}}}-\la \nu \ra _{X_{j,{\rm L}}\to X_{i,{\rm H}}} =0
}
for engines in the thermodynamic limit.
We here used the fact that an engine at the CE has zero power~\cite{SS16, SST16, ST17}, and thus the probability flow with $X_{i,{\rm H}}\to X_{j,{\rm L}}$ and that with $X_{j,{\rm L}}\to X_{i,{\rm H}}$ are balanced.
In contrast, by defining
\eq{
\bar{g}(n):=\frac{g(n)}{\sum_{n'}g(n')},
}
we obtain
\balign{
&\la n\ra _{X_{i,{\rm H}}\to X_{j,{\rm L}}}-\la n\ra _{X_{j,{\rm L}}\to X_{i,{\rm H}}} \nt \\
=&\sum _n\frac{n g(n)e^{\beta n\Di \mu}}{\sum_{n'}g(n') e^{\beta n'\Di \mu}}-\sum _n\frac{n g(n)}{\sum_{n'}g(n')} \nt \\
\geq&\sum _n\frac{n g(n)(1+\beta n\Di \mu)}{\sum_{n'}g(n') (1+\beta n'\Di \mu)}-\sum _n\frac{n g(n)}{\sum_{n'}g(n')} \nt \\
=&\frac{ \sum_nn^2\bar{g}(n)-\( \sum _n n \bar{g}(n)\) ^2 }{1+ \beta\Di \mu \sum_nn\bar{g}(n)}\beta\Di \mu \nt \\
\geq&0. \lb{main-gen}
}
In the fifth line, we used the fact that the numerator in the fourth line is the variance of $n$ with respect to $\bar{g}(n)$.
Recalling the condition of zero entropy production \eqref{zero-leak}, we find that two inequalities in \eref{main-gen} should be equality.
We first analyze the second inequality.
For the case with finite $\Di \mu$, the second inequality becomes equality if and only if $g(n)\propto \delta (n-n^*)$.
For the case of the linear response regime ($\Di \mu \to 0$), we take another approach.
The fluctuation-dissipation relation~\cite{Todabook} suggests that the tight-coupling condition is equivalent to $\langle \hJ_X^2\rangle^{\rm eq} \langle \hJ_Y^2\rangle^{\rm eq}=(\langle \hJ_X\hJ_Y\rangle ^{\rm eq})^2$, where we denoted by $\hJ_X$ and $\hJ_Y$ the particle current and another current corresponding to work extraction, respectively.
If $g(n)$ has finite fluctuation, this fluctuation can induce finite particle current $\hJ_X$ without inducing another current $\hJ_Y$, which violates the above tight-coupling condition.
Hence, $\la n\ra _{X_{i,{\rm H}}\to X_{j,{\rm L}}}-\la n\ra _{X_{j,{\rm L}}\to X_{i,{\rm H}}}=0$ holds if and only if $g(n)\propto \delta (n-n^*)$ both in and beyond the linear response regime.
We next consider the first inequality.
Under the above condition of $g(n)$, the first inequality becomes equality only when the peak of $g(n)e^{\beta\Di \mu n}$ is also at the same $n^*$.

Our findings are summarized as follows.
A stationary engine attains the CE only when $g(n)$ has delta-function-type singularity such that:
\balign{
P _{X_{i,{\rm H}}\to X_{j,{\rm L}}}(n )&\propto g(n)e^{\beta \Di \mu n}\propto \delta (n-n^*), \lb{sing-1} \\
P_{X_{j,{\rm L}}\to X_{i,{\rm H}}}(n )&\propto g(n)\propto \delta (n-n^*), \lb{sing-2}
}
for finite-size engines and
\balign{
P _{X_{i,{\rm H}}\to X_{j,{\rm L}}}(\nu )&\propto \lim_{V_0\to \infty} V_0 g(\nu V_0)e^{\beta \Di \mu \nu V_0}\propto \delta (\nu-\nu^*), \lb{sing-3} \\
P _{X_{j,{\rm L}}\to X_{i,{\rm H}}}(\nu )&\propto \lim_{V_0\to \infty} V_0 g(\nu V_0)\propto \delta (\nu-\nu^*), \lb{sing-4}
}
for engines in the thermodynamic limit.
Here, $P_{X\to X'}(n)$ represents the probability of the number of particles when the transition $X\to X'$ occurs, and $P_{X\to X'}(\nu )$ represents the probability density with respect to $\nu$.
The coefficient $V_0$ in Eqs.~\eqref{sing-3} and \eqref{sing-4} appears in order to translate the transition rate with $n(=\nu V_0)$ to the transition rate density with $\nu$.
It is easy to check that the above conditions are also sufficient for stationary engines to attain the CE.
The equations \eqref{sing-1}, \eqref{sing-2} (or \eqref{sing-3}, \eqref{sing-4}) require two conditions:
\be{
\renewcommand{\labelenumi}{(\arabic{enumi}).}
\item The probability $P_{X\to X'}(n)$ ($P_{X\to X'}(\nu)$) is a delta-function with respect to $n$ ($\nu$).
\item The two delta-functions have peaks at the same $n^*$ ($\nu ^*$).
}
To attain the CE, these two conditions should be satisfied for any transition between a state with H and that with L.
In the remainder of this paper, we treat only a single transition, and we say that the engine attains the CE if the analyzed transition satisfies the above two conditions.

\subsection{Consequence for finite-size engines}\lb{fin}

\figin{8.5cm}{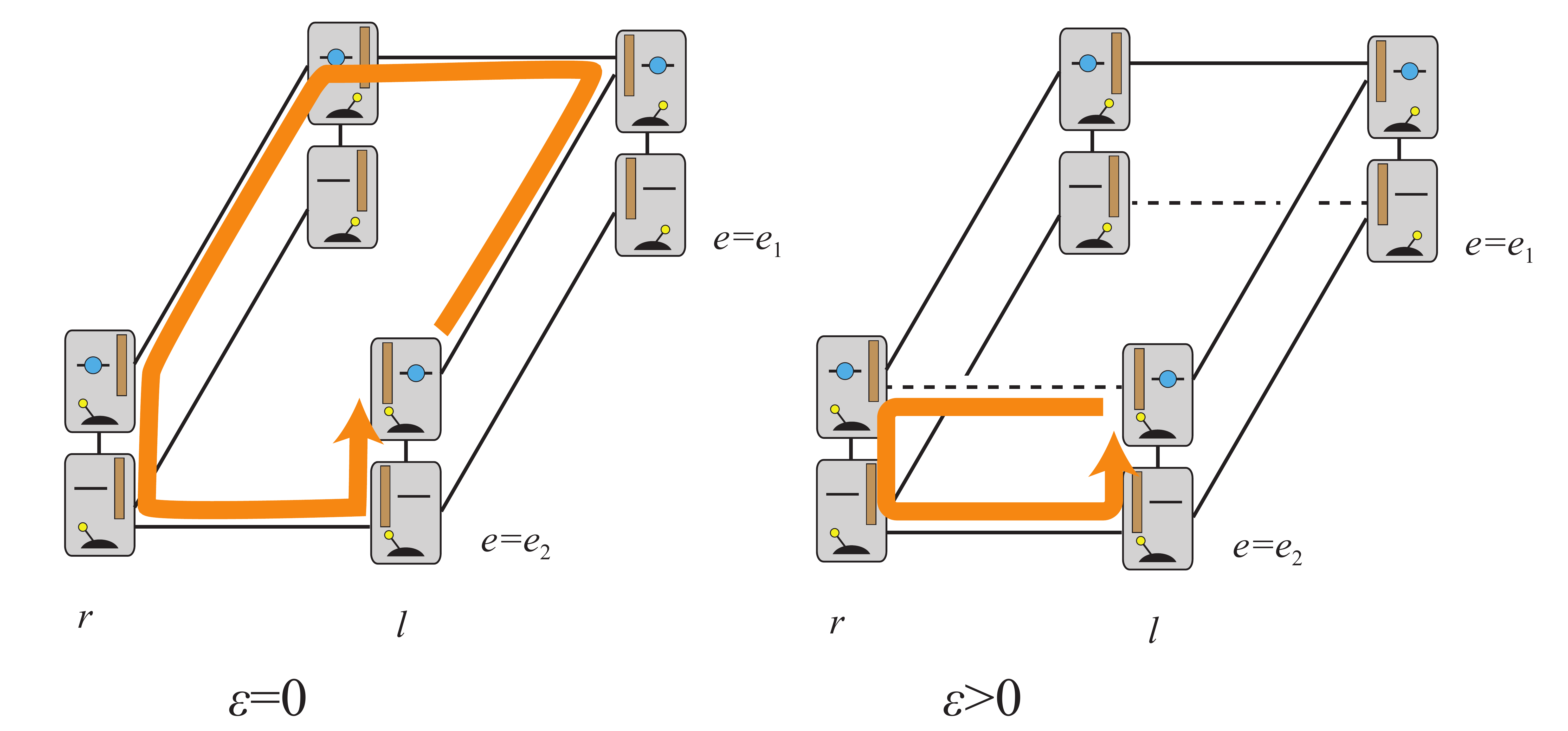}{
State space of the IES with $\ep=0$ (left) and $\ep>0$ (right) and their probability flow.
In the case of $\ep>0$, there exists particle flow without extracting work.
}{finite}

A finite-size engine does not generally satisfy the condition (1).
By contrast, the condition (1) accompanies the condition (2) because the presence of a transition ensures the existence of its opposite transition.
Thus, the nontrivial question is what types of engines satisfy the condition (1).
To realize the condition (1) in finite-size engines, the transition rate with $X_{i,{\rm H}}\to X_{j,{\rm L}}$ should be in the following form:
It takes a nonzero value only with a particular $n$, that is, $P_{X_{i,{\rm H}}\to X_{j,{\rm L}};n}=0$ for all $n\in \mathbb{N}$ except $n=n^*$.
In this setup, the particle current and the work current (or another particle current) are rigidly coupled and particle leakage is intrinsically prohibited.
Otherwise, particle leakage inevitably exists, which prohibits to attain the CE.
We remark that this condition is same for both cases in and beyond the linear response regime.
An important consequence of this finding is that if an engine is tight-coupling in the linear response regime, the engine always attains the CE beyond the linear response regime.

The IES introduced in Sec.~\ref{IES}, which is a prototype of finite-size engines, clearly exhibits the above findings.
Let us first consider the case with $\ep=0$, which corresponds to the tight-coupling condition.
In this case, the IES always attains the CE with any $\muH$ and $\muL$ even beyond the linear response regime by tuning $E^+-E^-=\muH -\muL$.
In contrast, the IES with $\ep>0$ never attains the CE even in the $\Di \mu \to 0$ limit.
In this case, there exists a cyclic path $(r,0,e_2)\to (r,1,e_2) \to (l,1,e_2) \to (l,0,e_2)\to (r,0,e_2)$, which transports a particle from the bath H to L without extracting work (see \fref{finite}).
This particle leakage prevents the IES to attain the CE even in the linear response regime.

\subsection{Consequence for macroscopic engines}\lb{macro}

The situation is completely different for the case of a macroscopic engine in the thermodynamic limit.
If the bath separation condition and equilibration condition are satisfied, the condition (1) is always satisfied due to the law of large numbers.
However, the condition (2) is not satisfied in general.
Thus, the nontrivial question is what types of engines satisfy the condition (2).

Using the Helmholtz free energy $F(V,n)$ and defining $r_{i,{\rm H}}:=V_{i,{\rm H}}/V_0$ with a typical volume of the engine $V_0$, we define a function $h(\nu )$ as
\eqa{
h(\nu):=\beta F\( r_{i,{\rm H}},\nu \) - \lim_{V_0\to \infty}\frac{1}{V_0} \ln V_0 P_{X_{i,{\rm H}}\to X_{j,{\rm L}};\nu V_0}.
}{def-h}
The probability density of $\nu$ under the condition that the transition $X_{i,{\rm H}}\to X_{j,{\rm L}}$ occurs is expressed in terms of $h(\nu )$ as
\balign{
\ln P_{X_{i,{\rm H}}\to X_{j,{\rm L}}}(\nu ) &= -h(\nu )+\beta \muH \nu +\const , \\
\ln P_{X_{j,{\rm L}}\to X_{i,{\rm H}}}(\nu ) &= -h(\nu )+\beta \muL \nu +\const .
}
Here, if $\del h/\del \nu$ has discontinuity at $\nu=\nu'$, we suppose that $\del h/\del \nu$ can take any value between $\lim_{\nu\to \nu'+0}\del h/\del \nu$ and $\lim_{\nu\to \nu'-0}\del h/\del \nu$.
Due to the law of large numbers, they have a delta-function type peak at $\nuH$ and $\nuL$, which are solutions of
\balign{
\ft{\frac{\del}{\del \nu}h(\nu)}{\nu =\nuH}=&\beta \muH ,\\
\ft{\frac{\del}{\del \nu}h(\nu)}{\nu =\nuL}=&\beta \muL .
}
We note that the second law of thermodynamics requires monotonic increase of $\del h(\nu)/\del \nu$ with respect to $\nu$, which confirms that the above equations have unique solutions.

\figin{5cm}{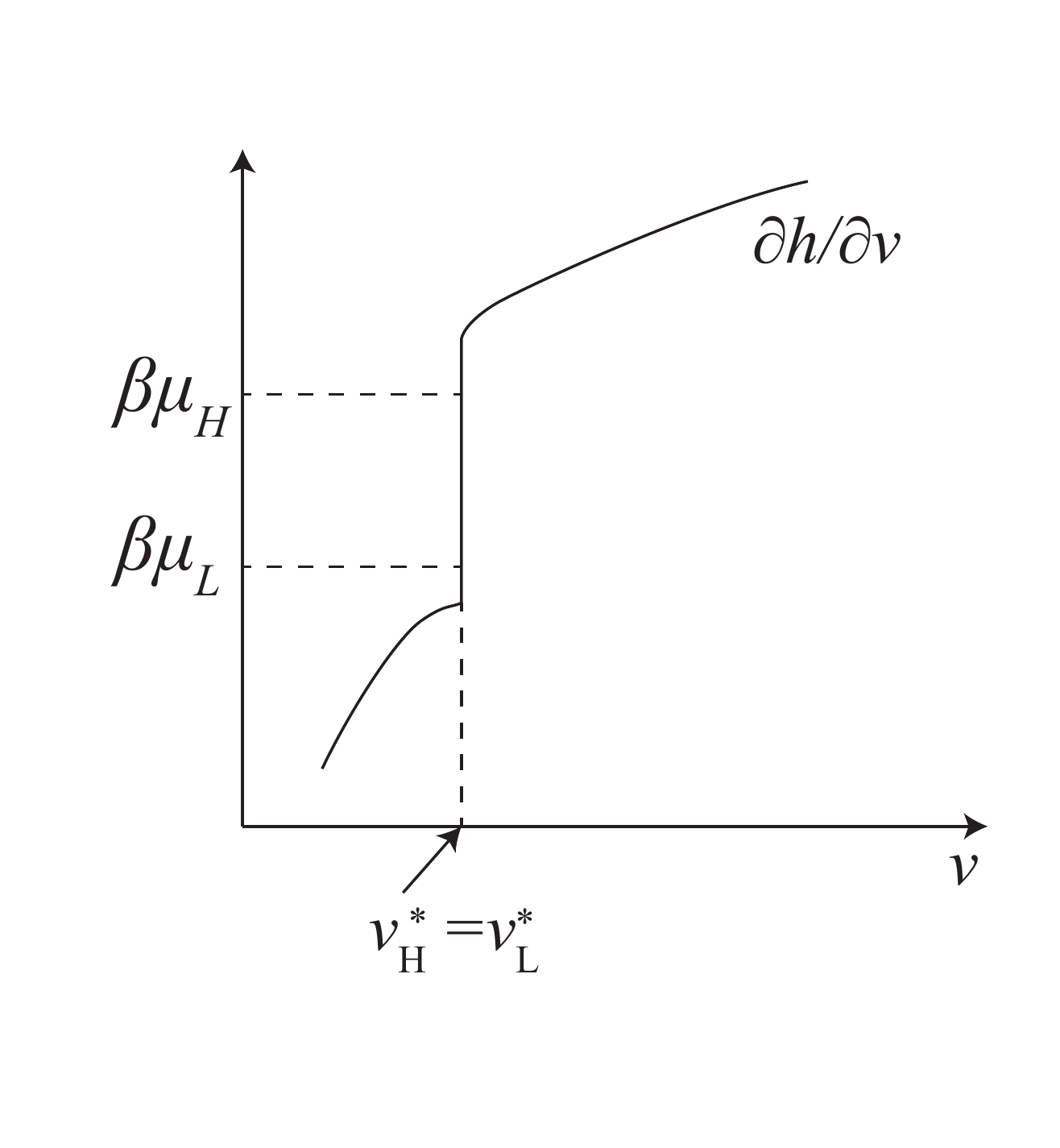}{
The behavior of $\del h(\nu)/\del \nu$ for an engine attaining the CE.
The indifferentiability of $h$ allows that two equations $\del h/\del \nu=\beta \muH$ and $\del h/\del \nu=\beta \muL$ have the same solution $\nu=\nu^*$.
}{dh-sing}

We draw many properties from these relations.
First, in the linear response limit $\muH-\muL \to 0$, the condition (2), $\nuH =\nuL$, is always satisfied.
Thus, if the bath separation condition and equilibration condition are satisfied, a macroscopic engine always attains the CE in the limit of $\Di \mu\to 0$.
One may feel that this result looks contradicting to experimental observations on macroscopic thermoelectric devices, where maximum efficiency is usually less than the CE.
However, our result and these observations are in fact consistent because general macroscopic thermoelectric devices do not satisfy the bath separation condition and equilibration condition in general.

Second, beyond the linear response regime $\muH \neq \muL$, the condition $\nuH=\nuL=:\nu^*$ reads
\balign{
\lim_{\nu' \to \nu^* -0}\ft{\frac{\del}{\del \nu}h(\nu)}{\nu =\nu'}\leq \beta \muL, \\
\lim_{\nu' \to \nu^* +0}\ft{\frac{\del}{\del \nu}h(\nu)}{\nu =\nu'}\geq \beta \muH .
}
These conditions are fulfilled in two cases.
In the first case, the engine has a discontinuous delta-function type transition rate.
We have already seen this idea for finite-size engines.
In this case, $\del h/\del \nu$ formally takes any real number at $\nu =\nu^*$, and the engine always attains the CE regardless of the amount of chemical potential difference.
In the second case, the engine has a continuous but indifferentiable transition rate or free energy whose derivative jumps at $\nu^* =\nuH=\nuL$.
In this case, $1/\beta \cdot \del h(\nu)/\del \nu$ has discontinuity, and an engine with two chemical potentials both of which are in this discontinuous interval always attains the CE.

The CGACE introduced in Sec.~\ref{CGACE}, which is a prototype of macroscopic engines, clearly illustrates these findings.
First, by setting $\muH-\muL \to 0$, the CGACE always attains the CE.
Second, for the case of $\Di \mu >0$, the CGACE attains the CE only when the transition rates between $A-D$ and $B-C$ are singular.
Otherwise, the CGACE fails to attain the CE.
These points are demonstrated in Appendix.~\ref{detail-CGACE}.

\subsection{Tight-coupling window}\lb{window}

To elucidate the difference of these various ways to achieve $\nuH=\nuL$, we introduce the idea of {\it tight-coupling window}.
Let $\mu_1$ and $\mu_2$ be given chemical potentials of two particle baths, and suppose that we can tune other parameters (e.g., external force, another pair of chemical potentials) to attain the CE.
We draw the $\mu_1 -\mu_2$ coordinate and plot pairs $(\mu_1, \mu_2)$ with which the CE is attainable (see \fref{window}).
An engine in the plotted region is tight-coupling in the sense that the particle flow and work flow are rigidly coupled with no leakage.
We call a rectangular plotted on the coordinate as {\it window}.

Our analyses in the previous sections suggest the possibility of three types of window:
\bi{
\item Infinitesimal window: The plotted region is only on $\mu_1=\mu_2$ and there is no window with finite width.
\item Finite window: A square with a edge from $\muL^*$ to $\muH^*$ is plotted.
\item Infinite window: The whole space of the coordinate is plotted.
}
Engines with the infinitesimal window attain the CE only in the linear response limit ($\Di \mu\to 0$), and those with other two windows attain the CE even beyond the linear response regime.
Our analyses discover that a finite-size engine has the infinite window or no window (i.e., the engine never attains the CE even in the linear response limit), while an engine in the thermodynamic limit may have all types of window.
A delta-function-type transition rate accompanies the infinite window.
In contrast, a continuous but indifferentiable transition rate or free energy accompanies the finite window.
The infinitesimal window is seen in general macroscopic engines with the bath separation condition and equilibration condition.
The CGACE with normal transition rates, the CGACE with singular transition rates, the IES with $\ep=0$ are examples of engines with the infinitesimal window, the finite window, and the infinite window, respectively.

\figin{8.5cm}{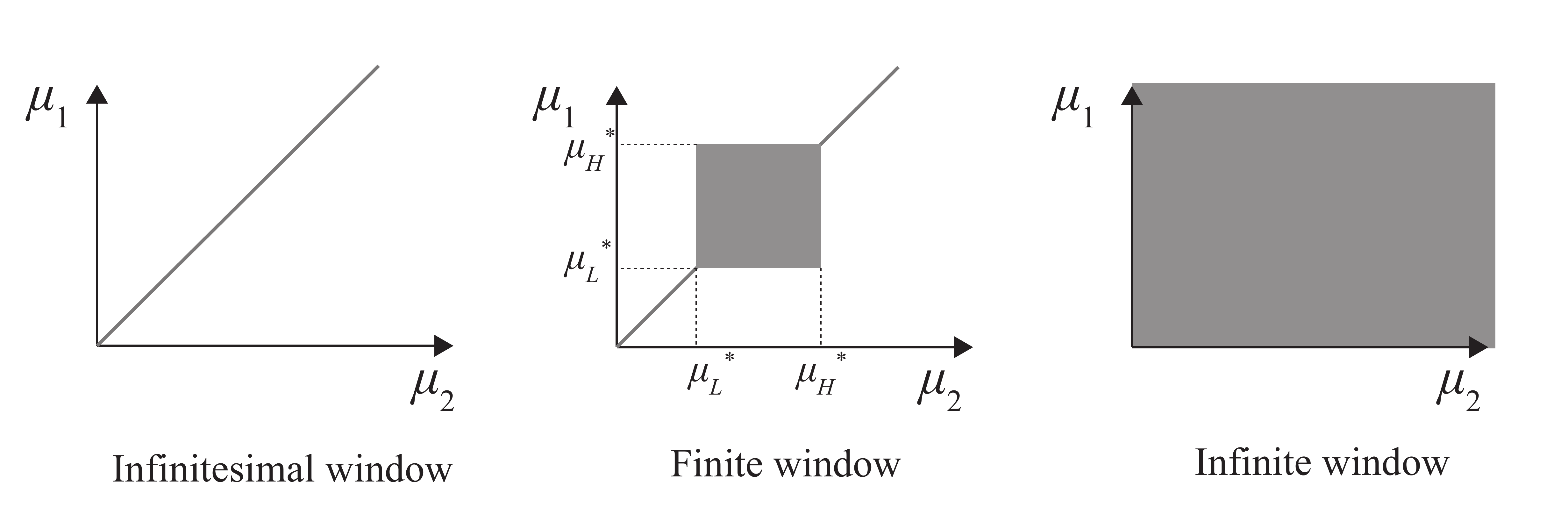}{
We plot pairs of chemical potentials with which the engine attains the CE on the $\mu_1-\mu_2$ coordinate with color gray.
We see three types of tight-coupling window.
}{window}

This viewpoint serves as a clear understanding of the connection between tight-coupling in the linear response regime and properties in the nonlinear regime.
It also characterizes the notion of tight-coupling beyond the linear response regime.
We remark that our characterization of macroscopic engines is on the foundation of microscopic understanding on transition rates.
This is similar to the case of the fluctuation-dissipation theorem, which connects the Onsager matrix and equilibrium fluctuation through microscopic analyses~\cite{Todabook}.

\section{Discussion}

We clarified the conditions for stationary engines to attain the CE, in particular the difference between finite-size engines and macroscopic engines.
We first derived the general condition Eqs.~\eqref{sing-1}-\eqref{sing-4} for stationary engines to attain the CE, and then examined its implications.
For finite-size engines, the attainment of the CE in the linear response regime directly leads to the attainment with any chemical potential differences.
In contrast, for macroscopic engines, there are three types of mechanisms to attain the CE, whose characteristics are elucidated by the tight-coupling window.
As long as satisfying the bath separation condition and equilibration condition, a macroscopic stationary engine attains the CE in the linear response limit.
To attain the CE beyond the linear response regime, an engine should employ discontinuous delta-function-type transition rate, or indifferentiable transition rate, or indifferentiable free energy.
An engine with the former one always attain the CE, while that with the latter two attains only when chemical potentials are within a certain definite range.

We remark that we have not used the detailed balance condition in the derivation of Eqs.~\eqref{sing-1}-\eqref{sing-4}, which suggests that our result is still valid for a system with broken time-reversal symmetry (e.g., a system with momentum or a magnetic field).
Our result relies on not a special type of symmetry but the zeroth law of thermodynamics.
This fact shows clear contrast to the result in Ref.~\cite{Shi15}, where the detailed-balance condition (time-reversal symmetry) is explicitly used in the derivation.

We briefly explain the origin of the difference between finite-size engines and macroscopic engines.
We see similarity between our result and the result on reversible adiabatic processes in small stochastic systems~\cite{STH, SSHT}.
Refs.~\cite{STH, SSHT} claims that the adiabatic process of a small system in a cyclic process is in general irreversible even in the quasistatic limit.
This shows clear contrast to the case of a macroscopic engine in the thermodynamic limit, where an adiabatic process in the quasistatic limit is reversible.
For both our result and the above fact on adiabatic processes, the difference between small systems and macroscopic systems lies in the following fact:
In the case of small systems quantities of $O(1)$ is taken into account, while in the case of macroscopic systems quantities of $o(V)$ is neglected.
This difference may be crucial when we consider thermodynamic properties, in particular thermodynamic reversibility.

\acknowledgements
This work is supported by Grant-in-Aid for JSPS Fellows Number 26-7602.

\

\appendix

\section{Properties of CGACE}\lb{detail-CGACE}

\subsection{Upper bound on efficiency}\lb{upper-CGACE}

We here calculate the upper bound on the efficiency of the CGACE $\eta :=W_{\rm tot}/C_\mu$, where $C_\mu$ represents the average consumption of chemical potential in a single rotation $A\to B\to C\to D\to A$.
To realize the probability current in the direction $A\to B\to C\to D\to A$, the following conditions should be satisfied
\balign{
j_{A\to D}(\nu_A^*)<&j_{D\to A}(\nu_D^*), \lb{jcond1} \\
j_{C\to B}(\nu_C^*)<&j_{B\to C}(\nu_B^*). \lb{jcond2}
}
Summing the logarithms of Eqs.~\eqref{jcond1} and \eqref{jcond2}, substituting \eref{jXX-} into them, and using the detailed balance condition \eqref{ldb}, we arrive at a key inequality:
\balign{
W_{\rm tot}
<&[F(r_A,\nu_A^*)-F(r_A,\nu_D^*)-F(r_B,\nu_B^*)+F(r_B,\nu_C^*) \nt \\
&+\muH (\nu_B^*-\nu_A^*)+\muL (\nu_D^*-\nu_C^*)]V_A \nt \\
&+\ln \frac{P_{A\to D,\nu_D^* V_A}P_{B\to C,\nu_B^*V_A}}{P_{A\to D,\nu_A^*V_A}P_{B\to C,\nu_C^*V_A}} \nt \\
=& \[ \( \muH \nu_B^*-\muL \nu_C^*-\frac{1}{\beta} \int _{\nu_C^*}^{\nu_B^*}\frac{\del h_B(\nu)}{\del \nu}d\nu\) \right. \nt \\
&\ \ \left. -\( \muH \nu_A^*-\muL \nu_D^*-\frac{1}{\beta} \int _{\nu_D^*}^{\nu_A^*}\frac{\del h_A(\nu)}{\del \nu}d\nu\) \] V_0 . \lb{key}
}
Here, we defined $h_X(\nu)$ ($X=A, B$) following \eref{def-h} as
\balign{
h_X(\nu)&:=\beta \( F(r_X,\nu)-\frac{1}{\beta V_0}\ln V_0 P_{X\to X^-;\nu V_0}\) 
}
with setting $V_0=V_A$.
Different from \eref{def-h}, we do not take $V_0\to \infty$ limit at this moment.
We also used relations
\balign{
\ft{\frac{\del h_A(\nu)}{\del \nu}}{\nu=\nu_A^*}=\ft{\frac{\del h_B(\nu)}{\del \nu}}{\nu=\nu_B^*}=&\beta\muH , \\
\ft{\frac{\del h_A(\nu)}{\del \nu}}{\nu=\nu_C^*}=\ft{\frac{\del h_B(\nu)}{\del \nu}}{\nu=\nu_D^*}=&\beta \muL ,
}
which follows from \eref{nX*}.
We note that the left-hand side in \eref{key} can approach arbitrarily close to the right-hand side.
The right-hand side of \eref{key} corresponds to the area of ``abcd" (colored by gray) in \fref{GX}.
In addition,  $C_\mu$ is evaluated as 
\eq{
C_\mu = (\muH -\muL )(\nu_B^*-\nu_D^*)V_0,
}
whose right-hand side corresponds to the area of ``pbqd" (surrounded by bold lines) in \fref{GX}.
As manifested in \fref{GX}, the efficiency is less than the CE with finite chemical potential difference $\Di \mu>0$ for normal setups.
In addition, it is also clear from \fref{GX} that the efficiency can reach the CE in the limit of $\Di \mu \to0$.

\figin{7cm}{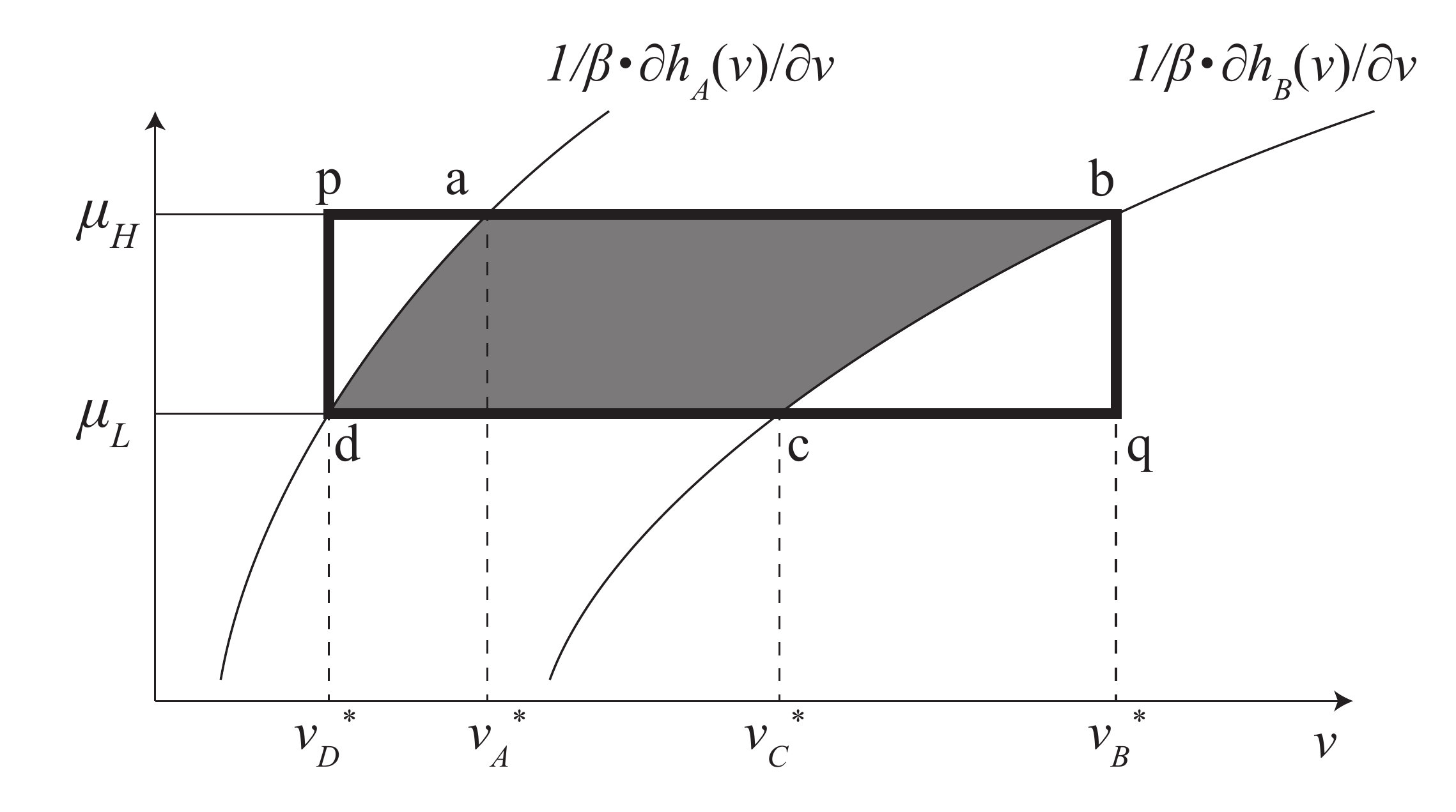}{
Graph of $1/\beta \cdot \del h_A(\nu)/\del \nu$ and $1/\beta \cdot \del h_B(\nu)/\del \nu$.
The area of gray region ``abcd" represents the upper bound of extracted work in one cycle, while the area of bold rectangular  ``pbqd" represents the consumed chemical potential in one cycle.
}{GX}

\subsection{Attainability to the Carnot efficiency}

To attain the CE with finite chemical potential difference, both $\del h_A(\nu)/\del \nu$ and $\del h_B(\nu)/\del \nu$ should jump from below $\muL$ to above $\muH$.
We here demonstrate how this condition is realized.
The following discussion is not rigorous in the point that we take the thermodynamic limit first and a priori neglect the existence of the fluctuation.
A rigorous justification from microscopic description is seen in Ref.~\cite{Shi15}.

We set the transition rates between $A$ and $D$ as 
\balign{
P_{A\to D;\nu V_0}&=\bcases{
k &:\nu \geq \bar{\nu }_{AD} \\
k\cdot e^{-\beta (F(r_D,\nu ) -F(r_A,\nu )-e_{DA})V_0} &: \nu < \bar{\nu }_{AD} 
}
\\
P_{D\to A;\nu V_0}&=\bcases{
k\cdot e^{-\beta (F(r_A,\nu )- F(r_D,\nu )+e_{DA})V_0} &:\nu \geq \bar{\nu }_{AD} \\
k&: \nu < \bar{\nu }_{AD} 
},
}
which can be realized with the Kawasaki-type transition rate.
Here, $\bar{\nu}_{AD}$ is the solution of
\eq{
F(r_A,\bar{\nu}_{AD})-F(r_D,\bar{\nu}_{AD})+e_{DA}=0.
}
The probability flow in terms of $\nu$ from $A$ to $D$ is then written as
\balign{
\ln j_{A\to D}(\nu )&=
\bcases{
 {\beta \( \muH \nu -F(r_A,\nu )\) }+c_l &:\nu \geq \bar{\nu }_{AD} ,\\
\beta (\muH \nu -F(r_D,\nu )+e_{DA})+c_l &: \nu < \bar{\nu }_{AD} ,
}
}
where $c_l:=\ln (kV_AP^{\rm ss}_{AB}/Z_{AB})$ is a constant.
$c_r$ is defined in a similar manner for $j_{D\to A}$.
In this setup, $\nu^*_A$ is given by the solution of
\eq{
\begin{array}{rll}
\muH=&\ft{\frac{\del F(r_A,\nu)}{\del \nu}}{\nu =\nu^*_A} &:{\rm for} \ f'_A < \muH  \\ 
\bar{\nu}_{AD}=&\nu^*_A &:{\rm for} \ f'_A\leq \muH \leq f'_D \\
\muH=&\ft{\frac{\del F(r_D,\nu)}{\del \nu}}{\nu =\nu^*_A} &:{\rm for} \ \muH < f'_D,
\end{array}
}
where we defined
\balign{
f'_A&:=\ft{\frac{\del F(r_A,\nu)}{\del \nu}}{\nu =\bar{\nu}_{AD}}, \\
f'_D&:=\ft{\frac{\del F(r_D,\nu)}{\del \nu}}{\nu =\bar{\nu}_{AD}}. 
}
In a similar manner, we find that $\nu^*_D$ is also given by the solution of
\eq{
\begin{array}{rll}
\muL=&\ft{\frac{\del F(r_A,\nu)}{\del \nu}}{\nu =\nu^*_D} &:{\rm for} \ f'_A< \muL  \\ 
\bar{\nu}_{AD}=&\nu^*_D &:{\rm for} \ f'_A \leq \muL \leq f'_D  \\
\muL=&\ft{\frac{\del F(r_D,\nu)}{\del \nu}}{\nu =\nu^*_D} &:{\rm for} \ \muL < f'_D.
\end{array}
}
Thus, $\nu^*_A=\nu^*_D$ holds if and only if
\eq{
f'_A\leq \muL \leq \muH \leq f'_D,
}
and here $\nu^*_A=\nu^*_D=\bar{\nu}_{AD}$ is satisfied.
In a similar manner, we find that $\nu^*_B=\nu^*_C$ holds if and only if  $\nu^*_B=\nu^*_C=\bar{\nu}_{BC}$.

The engine moves along $A\to B\to C\to D\to A$ if Eqs.~\eqref{jcond1} and \eqref{jcond2} are satisfied.
In this setup, the above two conditions read
\balign{
- (\muH -\muL ) \bar{\nu}_{AD} &>e_{DA}+c_l-c_r, \lb{nucond1} \\
(\muH -\muL ) \bar{\nu}_{BC} &>e_{AB}+e_{BC}+e_{CD}-c_l+c_r . \lb{nucond2}
}
Note that their summation
\balign{
\frac{C_\mu}{V_0}:=& (\muH -\muL ) (\bar{\nu}_{BC}- \bar{\nu}_{AD}) \nt \\
>&e_{AB}+e_{BC}+e_{CD}+e_{DA} \nt \\
=&\frac{W_{\rm tot}}{V_0}, \lb{nucond3}
}
is equivalent to the second law, and its right-hand side can reach arbitrarily close to the left-hand side by tuning $e_{AB}, \cdots, e_{AD}$.
In addition, using the balance condition $j_{B\to C}(\nu^*_B)=j_{D\to A}(\nu^*_D)$, $c_l-c_r$ in this setup is written as
\eq{
c_l-c_r = \muL \bar{\nu}_{AD} -\muH \bar{\nu}_{BC} +F(r_1,\bar{\nu}_{BC}) -F(r_D,\bar{\nu}_{AD})+e_{AB},
}
which implies that by tuning $e_{AB}$ properly both \eref{nucond1} and \eref{nucond2} can be simultaneously satisfied.
The above facts mean that the efficiency $\eta:=W_{\rm tot}/C_\mu$ can reach the CE.

\section{Additional parameter}\lb{add-para}

In Sec.~\ref{gen-pri}, we assumed that the energy difference of the engine between two states is independent of $n$.
However, several engines do not satisfy this assumption.
To treat such engines in our analyses, we here demonstrate the procedure to introducing an additional parameter if  $E_{XX^-}$ depends on $n$.
To say a result, the switch in the IES is a simple example of the additional parameter.
Therefore, we consider the conventional autonomous Maxwell's demon without switch, and demonstrate how the additional parameter (switch) is introduced.

The autonomous Maxwell's demon consists of two variables $(x,n)$, where $x\in \{ l,r\}$ represents the position of the wall and $n\in \{ 0,1\}$ represents the number of particle in the site.
The transition rates satisfy
\balign{
\ln \frac{P_{(r,0)\to (r,1)}}{P_{(r,1)\to (r,0)}}=&\beta\muH,  \\
\ln \frac{P_{(l,0)\to (l,1)}}{P_{(l,1)\to (l,0)}}=&\beta\muL,
}
and 
\balign{
\ln \frac{P_{(r,0)\to (l,0)}}{P_{(l,0)\to (r,0)}}=&\beta E^+,\\
\ln \frac{P_{(r,1)\to (l,1)}}{P_{(,1)l\to (r,1)}}=&\beta E^-.
}
The former two equations are the same as Eqs.~\eqref{4state1} and \eqref{4state2}, and the latter two are related to Eqs.~\eqref{4state3} and \eqref{4state4}.
In this setup,  $E_{XX^-}$ ($X$ takes $l$ or $r$) clearly depends on $n$.

We now introduce an additional parameter $e\in \{ e_1, e_2\}$, which is called switch in the main part.
We can see one-to-one correspondence between the above autonomous Maxwell's demon and the IES with $\ep=0$ and $c\to \infty$.
The addition of the switch separates the state $l$ ($r$) into two states $(l,e_1)$ and $(l,e_2)$ ($(r,e_1)$ and $(r,e_2)$), which makes $E_{XX^-}$ independent of $n$.

\end{document}